\documentstyle[12pt,epsf]{article} \hoffset -0.5in \textwidth 6.00in
\textheight
8.5in  \parskip 7pt \openup1.5\jot \parindent=0.5in
\newskip\humongous \humongous=0pt plus 1000pt minus 1000pt
\def\caja{\mathsurround=0pt}
\def\eqalign#1{\,\vcenter{\openup1\jot \caja
        \ialign{\strut \hfil$\displaystyle{##}$&$
        \displaystyle{{}##}$\hfil\crcr#1\crcr}}\,}
\newif\ifdtup

\def\eqright #1\cr{\noalign{\hfill$\displaystyle{{}#1}$}}
\def\eqleft #1\cr{\noalign{\noindent$\displaystyle{{}#1}$\hfill}}

\def\oldreffmt#1{\rlap{[#1]} \hbox to 2\parindent{}}

\def\figfmt#1{\rlap{Figure {#1}} \hbox to 1in{}}

\def\sectioneq{\def\theequation{\thesection.\arabic{equation}}{\let
\holdsection=\section\def\section{\setcounter{equation}{0}\holdsection}}}%

\newcounter{holdequation}

\def\auto{\eqno(\refstepcounter{equation}\theequation)}
\def\begineq #1\endeq{$$ \refstepcounter{equation}\eqalign{#1}\eqno
	(\theequation) $$}
\def\contlimit{\,{\hbox{$\longrightarrow$}\kern-1.8em\lower1ex
\hbox{${\scriptstyle (a\rightarrow0)}$}}\,}
\def\centeron#1#2{{\setbox0=\hbox{#1}\setbox1=\hbox{#2}\ifdim
\wd1>\wd0\kern.5\wd1\kern-.5\wd0\fi
\copy0\kern-.5\wd0\kern-.5\wd1\copy1\ifdim\wd0>\wd1
\kern.5\wd0\kern-.5\wd1\fi}}
\def\centerover#1#2{\centeron{#1}{\setbox0=\hbox{#1}\setbox
1=\hbox{#2}\raise\ht0\hbox{\raise\dp1\hbox{\copy1}}}}
\def\centerunder#1#2{\centeron{#1}{\setbox0=\hbox{#1}\setbox
1=\hbox{#2}\lower\dp0\hbox{\lower\ht1\hbox{\copy1}}}}
\def\lsim{\;\centeron{\raise.35ex\hbox{$<$}}{\lower.65ex\hbox
{$\sim$}}\;}
\def\gsim{\;\centeron{\raise.35ex\hbox{$>$}}{\lower.65ex\hbox
{$\sim$}}\;}

\def\super#1{\ifmmode \hbox{\textsuper{#1}}\else\textsuper{#1}\fi}
\def\textsuper#1{\newcount\holdspacefactor\holdspacefactor=\spacefactor
$^{#1}$\spacefactor=\holdspacefactor}

\def\getcite#1,{\advance\citenumber by1
\def\getcitearg{#1}\def\lastarg{@}
\ifnum\citenumber=1
\ref{#1}\let\next=\getcite\else\ifx\getcitearg\lastarg\let\next=\relax
\else ,\ref{#1}\let\next=\getcite\fi\fi\next}

\def\pom{{\rm P\kern -0.53em\llap I\,}}
\def\spom{{\rm P\kern -0.36em\llap \small I\,}}
\def\sspom{{\rm P\kern -0.33em\llap \footnotesize I\,}}

\relax

\newskip\humongous \humongous=0pt plus 1000pt minus 1000pt
\def\caja{\mathsurround=0pt}
\def\eqalign#1{\,\vcenter{\openup1\jot \caja
        \ialign{\strut \hfil$\displaystyle{##}$&$
        \displaystyle{{}##}$\hfil\crcr#1\crcr}}\,}
\newif\ifdtup

\def\eqright #1\cr{\noalign{\hfill$\displaystyle{{}#1}$}}
\def\eqleft #1\cr{\noalign{\noindent$\displaystyle{{}#1}$\hfill}}

\def\oldreffmt#1{\rlap{[#1]} \hbox to 2\parindent{}}

\def\figfmt#1{\rlap{Figure {#1}} \hbox to 1in{}}

\def\auto{\eqno(\refstepcounter{equation}\theequation)}
\def\begineq #1\endeq{$$ \refstepcounter{equation}\eqalign{#1}\eqno
	(\theequation) $$}
\def\contlimit{\,{\hbox{$\longrightarrow$}\kern-1.8em\lower1ex
\hbox{${\scriptstyle (a\rightarrow0)}$}}\,}
\def\centeron#1#2{{\setbox0=\hbox{#1}\setbox1=\hbox{#2}\ifdim
\wd1>\wd0\kern.5\wd1\kern-.5\wd0\fi
\copy0\kern-.5\wd0\kern-.5\wd1\copy1\ifdim\wd0>\wd1
\kern.5\wd0\kern-.5\wd1\fi}}
\def\centerover#1#2{\centeron{#1}{\setbox0=\hbox{#1}\setbox
1=\hbox{#2}\raise\ht0\hbox{\raise\dp1\hbox{\copy1}}}}
\def\centerunder#1#2{\centeron{#1}{\setbox0=\hbox{#1}\setbox
1=\hbox{#2}\lower\dp0\hbox{\lower\ht1\hbox{\copy1}}}}
\def\lsim{\;\centeron{\raise.35ex\hbox{$<$}}{\lower.65ex\hbox
{$\sim$}}\;}
\def\gsim{\;\centeron{\raise.35ex\hbox{$>$}}{\lower.65ex\hbox
{$\sim$}}\;}

\def\super#1{\ifmmode \hbox{\textsuper{#1}}\else\textsuper{#1}\fi}
\def\textsuper#1{\newcount\holdspacefactor\holdspacefactor=\spacefactor
$^{#1}$\spacefactor=\holdspacefactor}

\def\getcite#1,{\advance\citenumber by1
\ifnum\citenumber=1
\ref{#1}\let\next=\getcite\else\ifx#1@\let\next=\relax
\else ,\ref{#1}\let\next=\getcite\fi\fi\next}

\def\upon #1/#2 {{\textstyle{#1\over #2}}}
\relax
\renewcommand{\thefootnote}{\fnsymbol{footnote}}

\def\subhead#1{\bigskip\vbox{\noindent\bf #1}\nobreak\par}

\def\til#1{\centeron{\hbox{$#1$}}{\lower 2ex\hbox{$\char'176$}}}
\def\tild#1{\centeron{\hbox{$\,#1$}}{\lower 2.5ex\hbox{$\char'176$}}}
\def\sumtil{\centeron{\hbox{$\displaystyle\sum$}}{\lower
-1.5ex\hbox{$\widetilde{\phantom{xx}}$}}}

\def\pom{{\rm P\kern -0.53em\llap I\,}}
\def\spom{{\rm P\kern -0.36em\llap \small I\,}}
\def\sspom{{\rm P\kern -0.33em\llap \footnotesize I\,}}

\newcommand{\bit}{\begin{itemize}}
\newcommand{\eit}{\end{itemize}}

\newcommand{\beq}{\begin{equation}}
\newcommand{\eeq}{\end{equation}}
\newcommand{\beqa}{\begin{eqnarray}}
\newcommand{\eeqa}{\end{eqnarray}}

\begin{document}
\begin{titlepage}
\rightline{\vbox{\halign{&#\hfil\cr
&ANL-HEP-PR-94-84\cr
&\today\cr}}}
\vspace{0.25in}
\begin{center}

{\large\bf
THE SPECTRUM OF THE $O(g^4)$ SCALE-INVARIANT LIPATOV KERNEL}

\medskip

Claudio Corian\`{o} and Alan R. White
\footnote{Work supported by the U.S. Department of
Energy, Division of High Energy Physics, Contract\newline W-31-109-ENG-38}
\\ \smallskip
High Energy Physics Division, Argonne National Laboratory, Argonne, IL
60439.\\ \end{center}

\begin{abstract}

The scale-invariant $O(g^4)$ Lipatov kernel has been determined by t-channel
unitarity. The forward kernel responsible for parton evolution is evaluated
and its eigenvalue spectrum determined. In addition to a logarithmic
modification of the $O(g^2)$ kernel a distinct new kinematic component
appears. This component is infra-red finite without regularization and
has the holomorphic factorization property necessary for conformal
invariance. It gives a reduction (of ~$\sim 65\alpha_s^2/\pi^2\sim 0.15$) in
the power growth of parton distributions at small-x.

\end{abstract}

\renewcommand{\thefootnote}{\arabic{footnote}} \end{titlepage}

\subhead{1. INTRODUCTION}

The BFKL pomeron\cite{lip} or, more simply, the Lipatov pomeron, has
recently attracted growing attention, both from the theoretical and the
experimental side. The BFKL equation resums leading logarithms in $1/x$.
When applied in the forward direction, at large $Q^2$, it
becomes an evolution equation for parton distributions. The Lipatov pomeron
solution of the equation predicts that a growth of the form
$$ \eqalign {F_2(x,Q^2) ~\sim ~ x^{1-\alpha_0} ~\sim ~x^{-{1 \over 2}}} ~,
\auto\label{F2}
$$
where $\alpha_0 - 1$ is the leading eigenvalue of the forward $O(g^2)$
Lipatov kernel, should be observed in the small-x behaviour of structure
functions. The BFKL pomeron is important in hard diffractive processes in
general, for example deep-inelastic diffraction\cite{bar}, and, perhaps,
in rapidity-gap jet production\cite{ahm}. BFKL resummation is also
anticipated to play a key role in all semi-hard QCD processes
\cite{marchesini1}, where there is a direct coupling of the hard scattering
process to the pomeron. It is one of the major results of the HERA
experimental program that a growth similar to that of (\ref{F2}) is
observed\cite{der}.

{}From both a theoretical and an experimental viewpoint, it is vital to
understand how the BFKL equation, and (\ref{F2}) in particular, is affected
by next-to-leading logarithm contributions. In recent papers \cite{ker,ca}
the scale invariant part of the $O(g^4)$ next-to-leading kernel has been
determined by reggeon diagram and t-channel unitarity techniques. In this
paper we summarise some newly derived properties of this kernel,
concentrating on the forward direction relevant for the evolution of parton
distributions.

The new kernel is initially expressed in terms of transverse momentum
integrals. We have evaluated these integrals explicitly in the forward
direction. The results for the connected part of the kernel can be presented
in terms of finite combinations of logarithms. We find that there
are two components. The first simply has the structure of the $O(g^2)$ kernel
but with additional logarithms of all the transverse momenta involved. This
component can also be obtained by squaring the $O(g^2)$ kernel. The infra-red
divergences it produces after integration are regulated by the disconnected
part of the kernel. Also, for this component the new eigenvalues are
trivially obtained by squaring the $O(g^2)$ eigenvalues.

The second component is a new kinematic form which appears for the first
time at $O(g^4)$. It has a number of important properties. Firstly not only
is it separately finite, but it has no singularities generating infra-red
divergences after integration. It therefore requires no regulation. A
completely new eigenvalue spectrum is produced, which we give an explicit
expression for. We find that the spectrum posesses the fundamental property
of holomorphic factorization, which is a necessary condition for conformal
symmetry of the kernel\cite{lk}.

Since the new component appears first at $O(g^4)$ and also has the same
conformal invariance property as the leading-order kernel, we anticipate that
scale-ambiguities in its absolute evaluation will appear only in
higher-orders. That is to say it makes as much sense to evaluate this new
component at a fixed value of $\alpha_s$ as it did to evaluate the
leading-order contribution with such a value. Consequently we can quote a
result for the modification of $\alpha_0$ by this contribution. There is a
reduction of just the right order of magnitude to give an improvement in the
phenomenology, while preserving a significant effect.

We are unable, as yet, to give a complete result for how (\ref{F2}) is
modified by our results. This is because we must first determine how the
scale-invariance of the $O(g^4)$ kernel is broken by the off-shell
renormalization scale so that, presumably, $g^2 /4\pi \to \alpha_s(Q^2)$. This
is non-trivial since we expect that all the transverse momenta in the diagrams
of the kernel will be involved in the scale-breaking. Fadin and Lipatov have
already calculated\cite{fad} the full reggeon trajectory function, that
gives the disconnected piece of the kernel, in the next-to-leading log
approximation - including renormalization effects. The diagram structure we
have anticipated is what is found, but there are additional scale-breaking
internal logarithm factors involved. As outlined in \cite{ca}, we hope to
determine the scale-breaking logarithms, that occur in the remainder of the
kernel, by an extension of the Ward identity plus infra-red finiteness
analysis that gives the scale-invariant kernel.

The contribution of ($t$-channel) four-particle nonsense states to the
connected part of the $O(g^4)$ kernel is given in \cite{ker} as a sum of
transverse momentum integrals
$$
\eqalign{(g^2N)^{-2} K^{(4n)}_{2,2}(k_1&,k_2,k_3,k_4)_c~=~K_2~+~K_3~+K_4~}.
\auto\label{sum}
$$
To be consistent with the diagrammatic notation used below, we
introduce a momentum conserving $\delta$-function - compared to the definition
given in \cite{ker} - and write
$$
\eqalign{K_i~=~(2\pi)^3~\delta^2(k_1+k_2 - k_3 -k_4)~\tilde{K_i} , }
\auto\label{del}
$$
with
$$
\eqalign{\tilde{K_2}~=~- \sum_{\scriptscriptstyle 1<->2}
\Biggl({k_1^2J_1(k_1^2)k_2^2k_3^2+k_1^2J_1(k_1^2)k_2^2k_4^2+
k_1^2k_3^2J_1(k_3^2)k_4^2+k_1^2k_3^2k_4^2J_1(k_4^2) \over
(k_1-k_3)^2} \Biggr),}
\auto
$$

$$
\eqalign{\tilde{K_3}~=~\sum_{\scriptscriptstyle 1<->2}~
J_1((k_1-k_3)^2)\Bigl(k_2^2k_3^2+k_1^2k_4^2\Bigr),}
\auto
$$
and
$$
\eqalign{\tilde{K_4}~=~\sum_{\scriptscriptstyle 1<->2}~
k_1^2k_2^2k_3^2k_4^2~I(k_1,k_2,k_3,k_4), }
\auto
$$
where
$$
\eqalign{J_1(k^2)~=~{1 \over (2\pi)^3}\int d^2q {1 \over q^2(k-q)^2}}
\auto
$$
and
$$
\eqalign{ I(k_1,k_2,k_3,k_4)~=~{1 \over (2\pi)^3}\int d^2p {1 \over
p^2(p+k_1)^2(p+k_4)^2(p+k_1-k_3)^2}.}
\auto\label{box}
$$
The Ward Identity constraint that the kernel should vanish when $k_i \to
0,~i=1,..,4$,
together with infra-red finiteness, determine the relative weights of $K_2,
K_3$ and $K_4$.

It will be convenient to introduce a diagrammatic notation for transverse
momentum integrals. We define

\epsffile{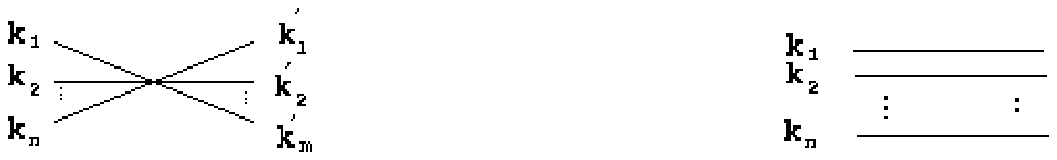}

\noindent $~=~~(2\pi)^3\delta^2(\sum k_i~  - \sum k_i')(\sum k_i~)^2~~~~
{}~~~~~~~~=~~(1/2\pi)^{3n}\int d^2k_1...d^2k_n~ /~k_1^2...k_n^2 .$

\vspace{.1in}

\noindent As we indicated above, we will define all kernels (and parts of
kernels) to include a factor $(2\pi)^3\delta^2(\sum k_i -\sum k_i') $. They
are then dimensionless and formally scale-invariant. $K_2, K_3$ and $K_4$
can be represented as a sum of diagrams of the form shown in Figs.~1(a), 1(b)
and 1(c) respectively.

\epsffile{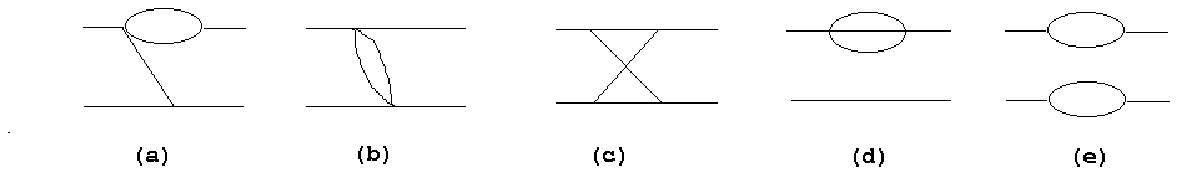}

\begin{description}

\item[Fig.~1] (a), (b), (c) - connected diagrams for the $O(g^4)$
kernel;~~ (d), (e) - disconnected diagrams.

\end{description}

In \cite{ker}, the disconnected part of the kernel was assumed to include
diagrams of the form of Fig.~1(d) only. Although generated by four-particle
nonsense states, diagrams of the form of Fig.~1(e) were not included. This
was essentially because they can not be associated with higher-order
reggeization. In fact such diagrams should be included in $K^{(4n)}_{2,2}$.
The point being that there is a further contribution to the $O(g^4)$ kernel,
from iteration of the two-particle nonsense state, which cancels the
contribution of such diagrams. Iteration of the two-particle nonsense state
gives a contribution of the form $[K^{(2)}_{2.2}]^2$, which can be
represented diagrammatically as in Fig.~2.

\vspace{.2in}

\epsffile{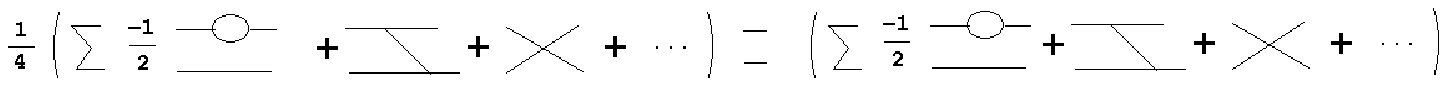}

\begin{description}

\item[Fig.~2] Iteration of the leading-order kernel via two-particle
nonsense states.

\end{description}

For $K^{(4n)}_{2,2}$ to be properly regulated after integration, diagrams of
the form of both Fig.~1(d) and Fig.~1(e) must be included and the result is
$$
\eqalign{(g^2N)^{-2} K^{(4n)}_{2,2}(k_1&,k_2,k_3,k_4)_c~=
{}~K_0~+~K_1~+~K_2~+~K_3~+K_4~,}
\auto\label{sum2}
$$
where $K_0$ contains the diagrams of the form of Fig.~1(e). If we use
(\ref{del}), $K_0$ is given by
$$
\eqalign{\tilde{K_0}~=~
\sum_{\scriptscriptstyle 1<->2}~
(2\pi)^3 J_1(k_1^2)J_1(k_2^2\Bigl(\delta^2(k_2-k_4)
+~\delta^2(k_2-k_3)\Bigr),}
\auto
$$
$K_1$ contains the diagrams of the form of Fig.~1(d) and
$$
\eqalign{\tilde{K_1}~=~-{2 \over 3}
\sum_{\scriptscriptstyle 1<->2}~
(2\pi)^3 k_1^2J_2(k_1^2)k_2^2\Bigl(k_3^2\delta^2(k_2-k_4)
+~k_4^2\delta^2(k_2-k_3)\Bigr),}
\auto
$$
with
$$
\eqalign{J_2(k^2)~=~{1 \over (2\pi)^3}\int d^2q {1 \over (k-q)^2}J_1(q^2).}
\auto
$$

That the $K_0$ contributions from $K^{(4n)}_{2,2}$ and $(K^{(2)}_{2,2})^2$
must cancel, determines that the full $O(g^4)$ kernel is given by
$$
\eqalign{ K^{(4)}_{2,2}~~=~~{1 \over 2^3}K^{(4n)}_{2,2} ~ -~(K^{(2)}_{2,2})^2}
\auto\label{full}
$$
This is the kernel that we wish to evaluate in the ``forward'' direction
$k_1=-k_2=k,~k_3=-k_4=k'$. Our result for $K^{(4)}_{2,2}(k,-k,k',-k')$ is
a much simpler expression than the full result given by (\ref{full}).

In writing down(\ref{full}) we have determined the overall sign by the
requirement that the contribution of the four-particle state should
be positive. The overall magnitude has been determined by noting that
the diagrams of the form of Fig.~1(e) contain only elements that appear in
the leading-order kernel and their contribution in $K^{(4n)}_{2,2}$ is
unambiguous. This implies that these diagrams should occur in
$K^{(4n)}_{2,2}$ (and therefore $\bigl(K^{(2)}_{2,2}\bigr)^2$) with an
absolute magnitude that is equal to that obtained by simple-minded iteration
 of the leading-order kernel.

The major technical problem in determining $K^{(4)}_{2,2}(k,-k,k',-k')$ is the
evaluation of the box graph, i.e. $I(k,-k,k',-k')$ defined in (\ref{box}).
As we will show in detail in \cite{ca2}, if we regularize $I$ with a mass
term $m^2$ in each propagator it can be evaluated as a sum of logarithms
associated with each of the possible ``two-particle'' thresholds in the
external momenta. As $m^2 \to 0$, we obtain
\beqa
&& 2\pi^2 I[k,k']= A_{12} Log[{ k'}^2/m^2] +
A_{23} Log[k^2/m^2] + A_{34} Log[{k'}^2/m^2]\nonumber \\
&& + A_{13} Log[(k+k')^2/m^2] +
A_{14}Log[k^2/m^2] +A_{24} Log[(k-k')^2/m^2]
\eeqa
where
$$
\eqalign{  &A_{12} = {k^2 - {k'}^2 \over k^2 (k+k')^2(k-k')^2}
{}~~~~~~~~~~~~~~~~~~~~~~A_{13}={1\over k^2 {k'}^2}\cr
&A_{14}={{k'}^2-k^2\over k'^2 (k+ k')^2 (k-k')^2}~~~~~~~~~~~~~~~~~~~~~
A_{23}={k'^2-k^2\over k'^2 (k+k')^2 (k-k')^2}\cr
&A_{24}={1\over k^2 {k'}^2}~~~~~~~~~~~~~~~~~~~~~~~~~~~~~~~~~~~~~~~~
A_{34}={k^2- {k'}^2\over k^2 (k+k')^2(k-k')^2}}
\auto
\label{f1}
$$
and so, as $m^2 \to 0$,
$$
\eqalign{
 \tilde{K}_4 ~\to ~ {-~k^2 {k'}^2 \over (2\pi^2)}&\Biggl(
{ 2({k'}^2 -k^2) \over (k+k')^2 (k-k')^2}
Log\Biggl[{{k'}^2\over k^2} \Biggr] \cr
&+ {1\over (k-k')^2} Log\Biggl[{ (k-k')^2\over m^2}\Biggr]
{}~+~{1\over (k+k')^2} Log\Biggl[{ (k +k')^2\over m^2}\Biggr]
\Biggr)~ .}
\auto\label{bm2}
$$
$\tilde{K}_3$ simply gives a contribution of the same form as the last
two terms in (\ref{bm2}), i.e. as $m^2 \to 0$
\beqa
\tilde{K}_3~\to~{k^2 {k'}^2 \over (2\pi^2)} \left(
{1\over (k-k')^2} Log\left[{ (k-k')^2\over m^2}\right]
+ { 1\over (k+k')^2} Log\left[{ (k+k')^2\over m^2}\right]\right)
\eeqa
Similarly $\tilde{K}_2$ gives
$$
\eqalign{
\tilde{K_2} \to {-k^2 {k'}^2 \over (2\pi^2) } & \Biggl(
{1 \over (k-k')^2} (Log\Biggl[{ k^2\over m^2}\Biggr] +
Log\Biggl[{ {k'}^2\over m^2}\Biggr] ) \cr
 + & { 1\over (k+k')^2} (Log\Biggl[ { k^2\over m^2} \Biggr] +
Log\Biggl[{ {k'}^2 \over m^2} \Biggr] ) \Biggr).}
\auto
$$

The infra-red finiteness of $\tilde{K}^{(4n)}_c ~=~ \tilde{K}_2 +
\tilde{K}_3 +\tilde{K}_4$ is now apparent and we can write
$$
\eqalign{2\pi^2\tilde{K}^{(4n)}_c~&=~
\Biggl( {k^2{k'}^2 \over (k-k')^2}Log\left[{(k-k')^4 \over k^2{k'}^2 }\right]
{}~+~ {k^2{k'}^2 \over (k+k')^2} Log\left[{(k+k')^4 \over k^2{k'}^2 }\right]
\Biggr)\cr
&~~~~~-~~~~
\Biggl( {2 k^2{k'}^2 (k^2 - {k'}^2) \over (k-k')^2(k+k')^2}Log\left[{k^2 \over
{k'}^2}\right] \Biggr) \cr
&=~~~~ \Biggl( ~~{\cal K}_1~~\Biggr) ~~-~~\Biggl( ~~{\cal K}_2~~\Biggr)
. }
\auto\label{4nc}
$$
Note that only ${\cal K}_1$ gives infra-red divergences (at $k'= \pm k$) when
integrated over $k'$. These divergences are cancelled by $\tilde{K}_0$ and
$\tilde{K}_1$. We will not discuss this cancellation explicitly, but
implicitly include $\tilde{K}_0$ and $\tilde{K}_1$ in ${\cal K}_1$ for the
rest of our discussion.

Apart from the logarithmic factors, ${\cal K}_1$ has the same structure as
the forward (connected) $O(g^2)$ kernel. Indeed, if we evaluate
all of the diagrams generated by Fig.~2 that survive in the forward
direction, it is straightforward to show that
$$
\eqalign{ {\cal K}_1 ~=~ (2\pi)^2 \widetilde{(K^{(2)}_{2,2})^2}~, }
\auto
$$
implying from (\ref{full}) and (\ref{4nc}) that
$$
\eqalign{ \tilde{K}^{(4)}_{2,2}~~=~~- {1 \over 2^4\pi^2} (3{\cal K}_1 ~ +~{\cal
K}_2)~.}
\auto\label{full2}
$$

There are a number of reasons to believe that the contribution of
$(K^{(2)}_{2,2})^2$ to the $O(g^4)$ kernel should produce the
scale-dependence of the $O(g^2)$ kernel. Indeed one might be tempted to
directly interpret the logarithms in ${\cal K}_1$ as associated with coupling
constant renormalization in the leading-order kernel. While this can not
be simply correct (real ulta-violet renormalization has to be involved to
bring in the correct asymptotic freedom coefficients) there may well be some
sense in which this is the case. We clearly have to carry out a full
scale-breaking analysis to determine what this sense may be.

The interesting part of (\ref{full2}) is the ${\cal K}_2$ component. This
is finite at $k=\pm k'$, and so does not generate any divergences when
integrated. The symmetry properties also determine that this term can only
appear at the first logarithmic level (since the antisymmetry of
$Log[k^2/{k'}^2]$ compensates for the antisymmetry of $(k^2-{k'}^2)$ ). It is
therefore a completely new feature of the $O(g^4)$ kernel.

We now move on to the eigenvalues of $K^{(4)}_{2,2}$. We use as a complete
set of orthogonal eigenfunctions
$$
\eqalign{ \phi_{\mu,n}(k')~=~({k'}^2)^{\mu}~e^{i n \theta}}~~~~~~~~\mu=~{1
\over2}
+i\nu,~~ n=0,\pm1,\pm2,...
\auto
$$
(Our definition of the kernel requires that we keep a factor of ${k'}^{-2}$ in
the measure of the completeness relation for eigenfunctions relative to
\cite{lip}). The eigenvalues of $(K^{(2)}_{2,2})^2$ are trivially given by
the square of the $O(g^2)$ eigenvalues, and so the essential problem is to
determine the eigenvalues of ${\cal K}_2$. As a preliminary we first define
${\cal K}_2$ for non-integer dimensions.

Since each logarithm in ${\cal K}_2$ originates from an integral of the form
of $J_1$ we can replace it by a simple integral of the form
\beqa
{k^2 \over 2\pi}\int {d^D q\over q^2 (k-q)^2}~=~\eta[k^2]^{D/2~-1}~,
\,\,\,\,\,\,\,\eta={\Gamma[2-D/2]\Gamma[D/2-1]^2\over \Gamma[D-2]},
\label{j1}
\eeqa
where $\eta \to 2(D-2)^{-1}$ when $D \to 2$. This gives
\beqa
{\cal K}_2~=~2\eta~ {k^2 {k'}^2 (k^2-{k'}^2)\over
(k+k')^2 (k-k')^2}\left( (k^2)^{D/2~-1} -
({k'}^2)^{D/2~-1}\right).
\label{cbox}
\eeqa
We now write
\beqa
 {\cal K}_2 \otimes \phi_{\mu,n} &=& {\cal K}_2^1 \otimes \phi_{\mu,n} -
{\cal K}_2^2 \otimes \phi_{\mu,n} \nonumber \\
&=& \lambda_1(\mu,n) \phi_{\mu,n} -
\lambda_2(\mu,n) \phi_{\mu,n} \nonumber \\
&=& \lambda(\mu,n) \phi_{\mu,n},
\eeqa
where
\beq
{\cal K}_2^1 \otimes \phi_{\mu,n}
=~2\eta \int {d^D k' \over ({k'}^2)^2} {(k^2)^{D/2} {k'}^2 (k^2-{k'}^2)
\phi_{\mu,n}(k')\over (k-k')^2(k+k')^2 ,}
\label{e1}
\eeq
and
\beq
{\cal K}_2^2 \otimes \phi_{\mu,n}
=~2\eta \int {d^D k' \over ({k'}^2)^2} { k^2 ({k'}^2)^{D/2} (k^2-{k'}^2)
\phi_{\mu,n}(k')\over (k-k')^2(k+k')^2 .}
\label{e2}
\eeq

We take the eigenfunction $\phi_{\mu,n}$ to be defined on a D-dimensional
angular space parameterized by $(\theta_1,\theta_2,...,\theta_{D-1})$
by assuming that $\theta\equiv \theta_{D-1}$. If we define
$cos\chi= k\cdot \hat{x}$ and $cos\theta ={k'}\cdot \hat{x}$, where $\hat{x}$
is an arbitrarily chosen unit vector, the only non-trivial angular integral
is
\beqa
I_{\chi}[n] &=& \int_{0}^{2 \pi} d\theta{ e^{i n \theta}
\over 1- z(k,k')sin^2\,\,(\theta- \chi)}
{}~~~~~~~~~z[{k,k'}]=-{4 k^2{k'}^2\over (k^2 -{k'}^2)^2} \nonumber \\
&=& \,2 \pi e^{i n \chi}
\left({k^2-{k'}^2\over k^2 + {k'}^2}\right)
\left[ \left( {k'\over k}\right)^n\Theta[k-k']
- \left({k\over k'}\right)^n \Theta[k'-k] \right].
\label{Ith}
\eeqa
if $n$ is an even integer ($\geq 0$). $I_{\chi}[n]$ vanishes if $n$ is an
odd integer and $I_{\chi}[-n] = I_{\chi}[|n|]$.

$I_{\chi}[n]$ is symmetric under the exchange of $k$ and $k'$, and also is
invariant under $k \to 1/k, k' \to 1/k'$. This last invariance is sufficient
to show from (\ref{e1}) and (\ref{e2}) that
$$
\eqalign{ \lambda(\mu~,n) ~=~\lambda(1 -\mu~,n)}
\auto\label{sym}
$$
Using (\ref{Ith}) we obtain from (\ref{e1}) and (\ref{e2}) that, as $D \to
2$,
\beqa
\lambda_1(\mu,n)~\to~2\eta
{\pi^{D/2}\over \Gamma[D/2]} \biggl(\beta\bigl(|n|/2 +D/2 +\mu - 1\bigr)
{}~-~\beta\bigl(|n|/2-D/2-\mu + 2\bigr) \biggr),
\eeqa
and
\beqa
\lambda_2(\mu,n)~\to~2\eta
{\pi^{D/2}\over \Gamma[D/2]} \biggl(\beta\bigl(|n|/2 +D +\mu - 2\bigr)
{}~-~ \beta\bigl(|n|/2-D -\mu + 3\bigr) \biggr),
\eeqa
where $\beta(x)$ is the incomplete beta function, i.e.
$$
\eqalign{
\beta(x)~&=~\int^1_0 dy~y^{x -1}[1+y]^{-1} \cr
&=~{1\over 2}\biggl(\psi\bigl({x+1\over 2}\bigr) -
\psi\bigl({x\over 2}\bigr)\biggr), \,\,\,\,\,\,
\psi (x)={d\over d\,\,x}log \Gamma[x]~.}
\auto
$$
$\lambda_1(\mu,n)$ and $\lambda_2(\mu,n)$ are separately singular at $D=2$,
but $\lambda(\mu,n)$ is finite, and writing $\Lambda(\nu,n)~\equiv~
\lambda( {1 \over 2} + i\nu~,n)$, we obtain
$$
\eqalign{ \Lambda(\nu,n) ~=~-~2\pi \biggl(\beta'\bigl({|n| + 1\over 2} +
i\nu\bigr)
{}~+~\beta'\bigl({|n| + 1 \over 2} -i\nu\bigr)\biggr). }
\auto\label{lam}
$$

We comment first on the general properties of (\ref{lam}). The symmetry
property (\ref{sym}) is clearly reflected in the presence of the two terms.
The two terms also give directly the property of holomorphic
factorization\cite{lk} necessary for conformal symmetry. That is
$\Lambda(\nu,n)$ is a sum of two terms, one depending on $(i\nu + 1/2 +
n/2)$ and the other on $( i\nu + 1/2 -n/2)$. These two combinations
determine respectively the eigenvalues of the holomorphic and
anti-holomorphic Casimir operators of linear conformal transformations.
Since

$$
\eqalign{
\beta'(x)~=~{1\over 4}\biggl(\psi'\bigl({x+1\over 2}\bigr) -
\psi'\bigl({x\over 2}\bigr)\biggr) , }
\auto\label{ps1}
$$
and
$$
\eqalign { \psi'(x)~=~\sum_{n=0}^{\infty} {1
\over (n+x)^2}, }
\auto\label{ps2}
$$
$\beta'(x)$ is a real analytic function and it follows from (\ref{lam}) that
the eigenvalues $\Lambda(\nu,n)$ are all real.

Note that since the eigenvalues of $K^{(2)}_{2,2}$ can be written as a sum of
holomorphic and antiholomorphic components it is clear that the eigenvalues
of $(K^{(2)}_{2,2})^2$ can not be. Therefore this part of the $O(g^4)$
kernel is not conformally invariant. This is one of the arguments, referred
to earlier, that this term is inter-related with the scale dependendence of
the $O(g^2)$ kernel.

Moving on to the modification of $\alpha_0$, we note that to obtain the
contribution to the eigenvalue of $\tilde{K}^{(4)}_{2,2}$ we multiply
$\Lambda(\nu,n)$ by $-1/~2^4\pi^2$. To compare with $\alpha_0-1$ we
have to multiply, in addition, by $N^2g^4/~(2\pi)^3$, where $N=3$ for QCD.
As we discussed above, since ${\cal K}_2$ represents a new kinematic form at
$O(g^4)$ we do not expect it to mix with renormalization effects and so it
should be legitimate to compare its contribution with $\alpha_0 -1$ by
setting $\alpha_s=g^2/4\pi$. It follows from the above that the leading
eigenvalue is $\Lambda(0,0)$, as it is for the $O(g^2)$ kernel. From
(\ref{lam})-(\ref{ps2}) we obtain the contribution to $\alpha_0 -1$ from the
${\cal K}_2$ term in (\ref{full2}) as
$$
\eqalign{ ~-~{ 9\alpha_s^2 \over 2\pi^3} \Lambda(0,0) ~
&= ~{18\alpha_s^2 \over \pi^2} \beta'(1/2)
= ~-~{9\alpha_s^2 \over 2\pi^2}\biggl(~\sum_{n=0}^{\infty} {1
\over (n+~1/4)^2} ~- ~\sum_{n=0}^{\infty} {1 \over (n+~3/4)^2}\biggr)\cr
&=~-~{9\alpha_s^2 \over 2\pi^2}\biggl(~16~+~{16 \over 25}~+~{16 \over 81}
{}~+~...~-~{16 \over 9}~-~{16 \over 49}~+~...\biggr)\cr
{}~&\sim ~-~{9\alpha_s^2 \over 2 \pi^2} \times 14.5
{}~\sim ~-~65~{\alpha_s^2 \over \pi^2} ~\sim ~-~0.15 }
\auto
$$
The corresponding contribution from the ${\cal K}_1$ term in (\ref{full2})
would be
$$
\eqalign{ - {3 \over 4 } \Biggl({12 \over \pi } Log[2] ~\alpha_s \Biggr)^2
{}~\sim~-~0.18}
\auto
$$
However, since we have no understanding how the logarithms in this term mix
with the renormalization of $\alpha_s$, this could well be essentially
accounted for by the choice of scale in the $O(\alpha_s)$ term. Therefore we
believe no attention should be paid to this last number.

\vspace{1in}

\centerline{\bf Acknowledgements}
We thank J. Bartels, R. K. Ellis, V. Fadin, R. Kirschner, L. Lipatov and
M. Wuesthoff for informative discussions and comments.

Finally, we dedicate this work to Roberto Baggio for the inspiration
provided. A mistake may be the final element leading to perfection, that
the midsummer dream of one of us did not come true notwithstanding.

\end{document}